\documentclass[useAMS,usenatbib]{mn2e}

\usepackage{graphicx}
\usepackage{amsmath}
\usepackage[bookmarks=false]{hyperref}

\def\aj{AJ}% Astronomical Journal
\def\apj{ApJ}% Astrophysical Journal
\def\apjl{ApJ}% Astrophysical Journal, Letters
\def\aap{A\&A}% Astronomy and Astrophysics
\def\baas{Bulletin of the American Astronomical Society}
\def\mnras{MNRAS}% Monthly Notices of the RAS
\def\nat{Nature}% Nature
\def\araa{ARA\&A}%
\def\procspie{Proc.~SPIE}%
\def\skytel{Sky\&Telescope}

\title[X-ray microlensing in the quadruply lensed quasar Q2237+0305]{X-ray microlensing in the quadruply lensed quasar Q2237+0305}
\author[F. Zimmer, R. W. Schmidt and  J. Wambsganss]{F.Zimmer$^{1}$\thanks{E-mail:
fzimmer@ari.uni-heidelberg.de}, R. W. Schmidt$^{1}$ and  J. Wambsganss$^{1}$\\
$^{1}$Astronomisches Rechen-Institut, Zentrum f\"ur Astronomie der Universit\"at Heidelberg, M\"onchhofstrasse 12-14, 69120 Heidelberg}
\begin{document}

\date{Accepted 13 December 2010; Received 10 November 2010; in original form 22 June 2010}

\pagerange{\pageref{firstpage}--\pageref{lastpage}} \pubyear{2010}

\maketitle

\label{firstpage}

\begin{abstract}
  We use archival data of NASA's Chandra X-ray telescope to compile an
  X-ray light curve of all four images of the quadruply lensed quasar
  Q2237+0305 ($z_{Q}$=1.695) from January 2006 to January 2007. We fit
  simulated point spread functions
  to the four individual quasar images using Cash's C-statistic to
  account for the Poisson nature of the X-ray signal.
  The quasar images display strong flux
  variations up to a factor of $\sim 4$ within one
  month.
  We can disentangle the intrinsic quasar variability from flux
  variations due to gravitational microlensing by looking at the
  flux ratios of the individual quasar images. Doing this, we find evidence
  for microlensing in image A.
  In particular, the time-sequence of the flux ratio A/B in the X-ray
  regime correlates with the corresponding sequence in the optical monitoring by OGLE in the
  V-band. The amplitudes in the X-ray light curve are larger. For the most prominent peak, the increase of the X-ray ratio
  A/B is larger by a factor $\sim 1.6$ compared to the signal in the
  optical. In agreement with theory and other observations of multiply
  imaged quasars, this suggests that the X-ray emission region of this
  quasar is significantly smaller than the optical emission region.
\end{abstract}


\section{Introduction}

The quasar Q2237+0305 was discovered in 1984 \citep{Huchra} during a
spectroscopic survey of nearby galaxies. The spectrum of the nucleus
of the barred spiral galaxy 2237+0305 ($z_{G}$=0.0394) was found to be
superimposed by a quasar component at a redshift of $z_{Q}$ =
1.695.\\ The first high resolution observations of the system showed
three images of the quasar \citep{tyson}. This number was soon
corrected by \citet{Yee} who first observed all four known point like
quasar images around the core of the galaxy. The images are arranged
in a nearly symmetric way, hence the name `The Einstein
Cross'.\\ Spectroscopy proved that they are images of a single quasar
\citep{lensmodel_schneider, 1989A&A...208L..15A}. The images are
separated by up to 1.8\arcsec\; from each other, and are labeled A
through D.\\ After the prediction by \citet{chang} and initial
theoretical studies (e.g., \citet{kayser, schneiderweiss, pacz}),
microlensing was first detected in Q2237+0305 \citep{irwin,
  wampaczschneid}. In fact, this detection of `quasar microlensing'
was the first evidence for microlensing generally, including stellar
microlensing in the galaxy which was not discovered until 1993
\citep{aubourg, alcock, udal93}. Today `The Einstein Cross' is one of
the best studied multiply lensed quasars and there have been a lot of
monitoring programs in the optical \citep{corrigan, monitoring_pen,
  ostensen, OGLE1, OGLE2, monitoring_APO, monitoring_alcade, OGLE3}
where much microlensing activity has been observed. The quasar has
also been detected in the UV \citep{Blanton}, the NIR \citep{AgolNIR},
the MIR \citep{AgolMIR} and the radio regime \citep{Falco}. However,
there is no published light curve of Q2237+0305 in the X-ray regime
yet.\\ The X-ray emission of Q2237+0305 was first detected with
\textit{ROSAT/HRI} observations in 1997 \citep{joachimROSAT}. Since
then there have been other X-ray observations of Q2237+0305 like a
single spectroscopic observation with \textit{XMM-Newton} from 2002
(data set ID: 0110960101; PI: Watson) \citep{fedorova} and two
\textit{Chandra} observations \citep{photonindex}. In this paper we
study ten archival \textit{Chandra} observations ranging from January
2006 until January 2007 and compile the first X-ray light curve of
Q2237+0305.

\section{Chandra X-ray data}

\begin{figure*}
	\begin{center}
        \includegraphics[bb=14 14 802 405, scale=.5]{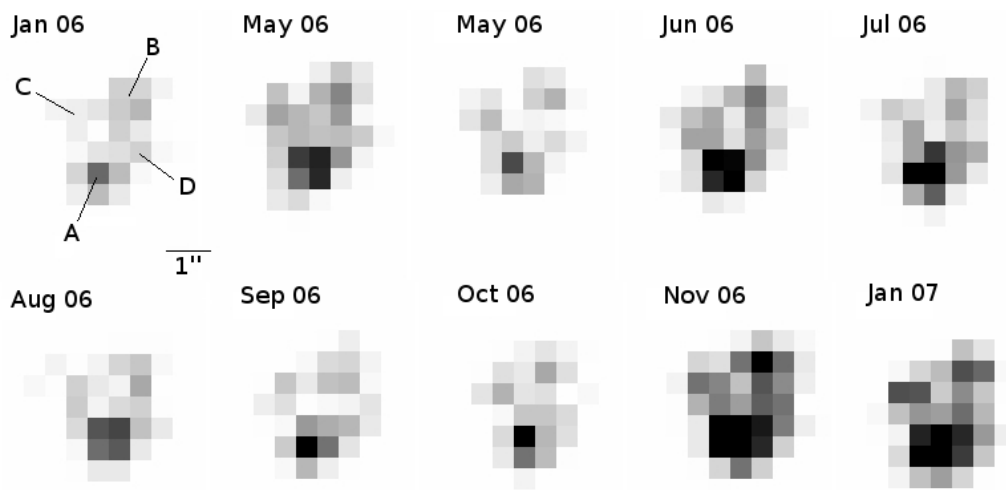}
        \end{center}
        \caption[X-ray images of Q2237+0305]{All ten \textit{Chandra}
          images of Q2237+0305 after recalibration and in the energy
          range between 0.5 keV and 8 keV as used here. Each image has
          its observation ID attached in the upper left. The brightest
          pixel of image A in January 2006 has 33 counts while the
          brightest pixel in image C has 6 counts. The greyscale is
          identical for each observation. Brightness variations in
          image A are clearly visible.}
\label{6831_raw_labled}
\end{figure*}

The data we use was taken by the \textit{Chandra X-ray Observatory}
\citep{weisskopf} in the period between January 9, 2006 (Julian date
2453745.0) and January 14, 2007 (Julian date 2454114.9)(PI:
Kochanek). The exposure time was $\sim$ 8 ks per observation. In Table
\ref{datatable} we list the ten observation IDs, the exposure times
and observation dates. The observation IDs are listed in order to
reference the individual images. We obtained the data from the Chandra
Data Archive. Figure \ref{6831_raw_labled} shows all ten
observations. The brightness variabilities of the four images are
already visible.  The four quasar images are labeled in the common way
from A to D.

\subsection{Chandra configuration}

All ten images were obtained with the ACIS-S detector in
\textit{VFAINT} mode. ACIS-S is one of the two detectors in the ACIS
(Advanced CCD Imaging Spectrometer) \citep{acis} instrument.  In our
data set Q2237+0305 was placed on the back-illuminated chip S3 (1024
$\times$ 1024 pixel).\\ The \textit{VFAINT} mode provides the observer
with the event position in detector coordinates, the event amplitude,
the arrival time, and the pixel values in a 5$\times$5 island
(compared to the 3$\times$3 island in \textit{FAINT} mode).  In our
case a subarray of only half of the ACIS-S array was used for the
benefit of reducing the read-out time to 1.7 seconds avoiding pile-up
effects.

\begin{table*}
   \centering
   \begin{minipage}{140mm}
   \caption[Exposure times and observation dates of the Chandra
     data]{The exposure times and starting dates of the Chandra
     observations of Q2237+0305 used here. The observation IDs are
     listed for future reference.\\}
    \label{datatable}
    \centering
    \begin{tabular}{@{}ccc@{}}
      \hline
      \textbf{Observation ID} & \textbf{Exposure Time/\textit{s}} & \textbf{Start Date} \\
      \hline
      6831 & 7263.84 & 2006-01-09 23:39:56 \\
      6832 & 7936.11 & 2006-05-01 00:46:10 \\
      6833 & 7952.71 & 2006-05-27 09:45:50 \\
      6834 & 7937.77 & 2006-06-25 17:24:13 \\
      6835 & 7871.38 & 2006-07-21 12:03:49 \\
      6836 & 7927.81 & 2006-08-17 03:59:06 \\
      6837 & 7944.41 & 2006-09-16 04:50:17 \\
      6838 & 7984.25 & 2006-10-09 03:41:03 \\
      6839 & 7873.04 & 2006-11-29 20:08:05 \\
      6840 & 7975.95 & 2007-01-14 22:30:10 \\
      \hline
   \end{tabular}
   \end{minipage}
\end{table*}

\subsection{Data processing}

To process the data we use the \textit{CIAO} \citep{ciao} software
package (version 4.2). All ten images are calibrated according to the
latest Calibration Database (CALDB, version 4.2.0) offered by the
\textit{Chandra X-ray Center}. We reprocess the images using the
standard level 1 event lists produced by the \textit{Chandra} pipeline
processing, including the latest gain maps and calibration products to
create new level 2 event files. However, we do not apply the
\textit{VFAINT} background correction but only the \textit{FAINT}
correction as the photons marked as background events in the
\textit{VFAINT} correction clearly clump arround the source location,
indicating that in the 5$\times$5 island, real events of the closely
separated four quasar PSFs have erroneously been flagged as
background. In any case, this affected only about ten
photons.\\ Simulations (PIMMS v3.9) show that for the brightest image
(image A) in the brightest exposure (6839), (see Figure
\ref{6831_raw_labled}) pile-up was only 4\% using the spectrum by
\citet{photonindex}.\\ We also check for contaminating flares during
the exposure of the single images and find that all ten images show
flat light curves. We check the light curves for both, source photons
(photons in the energy range between 0.5 keV and 8 keV) as well as
background events (photon energies higher than 9.5 keV).  Finally, we
consider an energy range from 0.5 keV to 8 keV in this paper. The
event files are converted into images with a pixel size of
0.492\arcsec.

\section{Photometry}

\subsection{The Chandra PSF}

For the purpose of doing photometry on each of the four images we
created a suitable point spread function (PSF) in order to later use
it for PSF-fitting. To account for Chandra's special Wolter optics
\citep{wolter} that is comprised in the \textit{High Resolution Mirror
  Assembly} (HRMA) \citep{weisskopf, schwartz, gaetz} and its special
imaging characteristics as well as for the blended nature of the four
quasar images (maximum separation of 1.8\arcsec) that requires a very
precise spatial resolution, we create a special PSF.  We use the two
tools especially made for this scope: \textit{ChaRT} (Chandra Ray
Tracer) \citep{chart}, a web interface to a raytrace code developed by
the \textit{Chandra X-ray Center} and MARX (Model of AXAF Response to
X-rays, AXAF was Chandra's name close before launch)
\citep{marx}. \textit{ChaRT} allows to simulate \textit{HRMA} PSFs at
any off-axis angle and for any energy or spectrum.

\subsubsection{Creating the PSF}

We use \textit{ChaRT} to simulate three monochromatic PSFs for the off-axis position
of Q2237+0305 each with the maximum ray density available 
in order to get a good signal-to-noise ratio. We do not simulate pile-up effects.
The photon energies of the PSFs are 0.5 keV, 2 keV and 8 keV. We choose the
energies with respect to the spectral range and their relative contribution
to the spectrum of Q2237+0305, which will be explained more detailed below.
The output of \textit{ChaRT} is processed with \textit{MARX} which projects
the rays onto the detector taking into account any detector effects.
In addition to simulating the detector response, \textit{MARX} uses the ray
weights to account for the mirror effects, i.e., different efficiencies of
different shells at different angles and energies.

\subsubsection{Combining the PSFs}

For the PSF fitting we calculate the sum of the three PSFs weighting each
energy according to its fraction in the spectra of Q2237+0305 as it is
seen by \textit{Chandra}. The weight of the single PSFs is estimated 
using \textit{XSPEC} \citep{xspec} for a photon index of $\Gamma=1.90^{+0.05}_{-0.05}$
\citep{photonindex} and a galactic absorption of \mbox{$N_{H}=5.5\times10^{20}$ cm$^{-2}$}
\citep{dickey}. The resulting relative weights of the PSFs are 1.000 for the
0.5keV PSF, 0.103 for the 2keV PSF and 0.008 for the 8keV PSF, respectively.

The final PSF is dominated by the 0.5 keV PSF, however, it has
slightly broader wings. It is subsampled by a factor of two
(corresponding to a binning of $\sim 0.25\arcsec \mathrm{per\;pixel}$)
and trimmed to a size of $\sim25\arcsec\times25\arcsec$ (102 px
$\times$ 102 px) to facilitate the shifting of the PSF. We use this
PSF to construct the quadruply PSF configuration of our images.

\subsection{PSF fitting}

As the four images of Q2237+0305 are very close to each other (within
a circle of 1.8\arcsec diameter), their PSFs are blended and we need a
sophisticated program to disentangle and isolate the individual
fluxes. Thus, we choose GALFIT 2.0.3 \citep{galfit}. GALFIT is a
two-dimensional fitting program originally designed to extract
structural components from galaxy images. Nevertheless, as GALFIT is
able to fit several components simultaneously, i.e., quasar images, it
is an appropriate choice in particular because it is able to fit
user-provided PSFs. Compared to counting photons in circular apertures
we can measure the flux of the whole PSF and obtain very precise
limits of the order of 0.05\arcsec on the position of the PSF on the
quasar images. In our case, we fit four user-provided PSFs
simultaneously in order to get the fluxes of each individual image A,
B, C and D. The four PSFs have fixed relative positions while the
absolute position of the template of the four PSFs is fitted. Here,
the absolute position is the origin of the relative PSF positions
listed in Table \ref{q2237astrometric}, i.e. the position of quasar
image A. We use the relative positions obtained by \citet{Blanton}
from UV data because these are minimally affected by galaxy
light. \citet{Blanton} determined the astrometric properties of the
lens system with the \textit{HST WFPC2} camera in the UV (F336W and
F300W filters). We also adapt GALFIT for the X-ray regime to fit
fluxes instead of magnitudes.

\begin{table}
\centering
\caption[Relative positions of the four images of
  Q2237+0305]{Astrometry of the quadruply imaged quasar Q2237+0305:
  The table lists the relative positions of the four quasar images as
  presented by \citet{Blanton}\\}
\label{q2237astrometric}
\begin{tabular}{@{}ccc@{}}
\hline
   \textbf{Component} & \textbf{Right Ascension/\arcsec} & \textbf{Declination/\arcsec}\\
\hline
A & $0.000\pm0.0015$ & $0.000\pm0.0015$ \\
B & $-0.671\pm0.0015$ & $1.697\pm0.0015$ \\
C & $0.634\pm0.0015$ & $1.210\pm0.0015$ \\
D & $-0.866\pm0.0015$ & $0.528\pm0.0015$ \\
\hline
\end{tabular}
\end{table}

\subsubsection{The fitting algorithm}

The fitting algorithm used in GALFIT is based on the
\textit{Levenberg-Marquardt} algorithm \citep{numericalrecipes,
  Bevington} that provides a numerical solution to the problem of
minimising a function over a set of nonlinear parameters of the
function.  By default, GALFIT and its implemented
\textit{Levenberg-Marquardt} algorithm is using a least-squares
minimization that is based on the assumption of normally distributed
photon counts. However, to cope with the purely Poissonian nature of
the X-ray signal we fit the data by using a statistic that is based on
a Poisson distribution. For this we adapt GALFIT to minimize Cash's
C-statistic \citep{Cash} (see Appendix) instead of the original
$\chi^{2}$-statistic. Indeed, \citet{Humphrey} demonstrate that
$\chi^{2}$-statistics can give rise to intrinsically biased parameter
estimates and that the use of Cash's C-statistic gives comparatively
unbiased parameter estimates. In our case we choose a slightly
modified form as implemented in XSPEC \citep{xspec}\footnote{Please
  refer to the XSPEC manual for a more detailed derivation.}:

\begin{equation}
   \begin{split}
      C=-2\cdot\displaystyle\sum_{i=1}^{N}&y_{i}\cdot \log(\tilde{y_{i}}(z,a_{1},..,a_{M}))-\tilde{y_{i}}(z,a_{1},..,a_{M})\\
      &+y_{i}-y_{i}\cdot \log(y_{i})
    \end{split}
\label{Cequation}
\end{equation}

Here $y_{i}$ is the $i$-th datapoint that is fitted to the model
$\tilde{y_{i}}(z,a_{1},..,a_{M}))$ with $M$ adjustable parameters
$a_{j}$. The C-value is an indicator for the goodness-of-fit and is
smallest for the best-fitting model.

\subsubsection{The fitting method}

\begin{figure*}
	\begin{center}
        \includegraphics[scale=0.5]{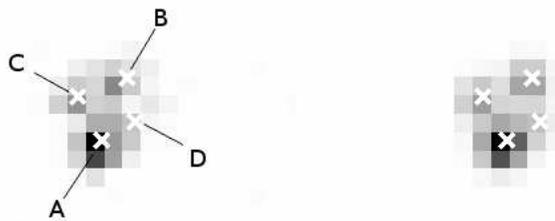}
        \end{center}
\begin{minipage}{140mm}
\caption[Comparison between the central 8$\times$8 pixels in raw
  Chandra image of Q2237+0305 in the 0.5 keV - 8keV band and its best
  fitting model]{Comparison between the central 8$\times$8 pixels in
  the raw Chandra image (6838) of Q2237+0305 in the 0.5 keV - 8keV
  (left panel) and its best fitting model (right panel). The greyscale
  for both images is identical. The crosses mark the positions of the
  quasar images as obtained in the best fit.}
\label{vorher_nachher}
\end{minipage}
\end{figure*}

To fit the four PSFs and thereby estimate the fluxes of the four
images, it is sufficient to choose relatively coarse estimates for the
initital fluxes and the position.  GALFIT then typically converges
after a few iterations. It determines the PSF template position with
an accuracy of $\frac{1}{10}$ of a pixel (0.05\arcsec). The
convergence is robust against changes in the initial conditions.
Nevertheless, we confirm the fitting results by manually stepping
through the six-dimensional (four fluxes and two coordinates)
parameter space.\\ Once the best fit is found, we use the fluxes
provided by GALFIT. The method was carried out on all ten
\textit{Chandra} images. Figure \ref{vorher_nachher} shows the raw
Chandra image (ID 6838) and its best-fitting model. It is notable how
well the Einstein Cross is reproduced and how good the model reflects
the relative brightness of the four quasar images compared to the raw
image.

To estimate the error on the fluxes we manually determine the 68\%
contours for each quasar image flux in each epoch using the $\Delta
C=1$ contour with respect to the minimum C-value found for the
best-fit.

\subsection{Robustness of the fit}

With the implementation of Cash's C-statistic in GALFIT we found an
ideal way of fitting the X-ray data. The nature of our signal is
clearly Poissonian and this is reflected in Cash's statistic. For
example, the total number of counts for observation 6831 is less than
300 with 33 counts in the brightest pixel.\\ One of the crucial facts
is that the Poisson distribution is discrete since the counts in one
pixel can only be a natural number. The continuous normal distribution
does not have this limitation, furthermore the normal distribution
even allows negative values, an aspect that clearly rules out
$\chi^{2}$-based fitting of low number counts. The asymmetric Poisson
distribution assures that the fitting takes better care of pixels with
a low number of counts ($<3\;\mathrm{counts}$) and therefore provides
an accurate description of the pixels. This is very important because
all of the quasar image PSFs have pixel counts in exactly that count
regime.

\section{Results}

\begin{figure*}
  \begin{center}
\begin{minipage}{170mm}
        \centering
	\includegraphics[scale=0.8]{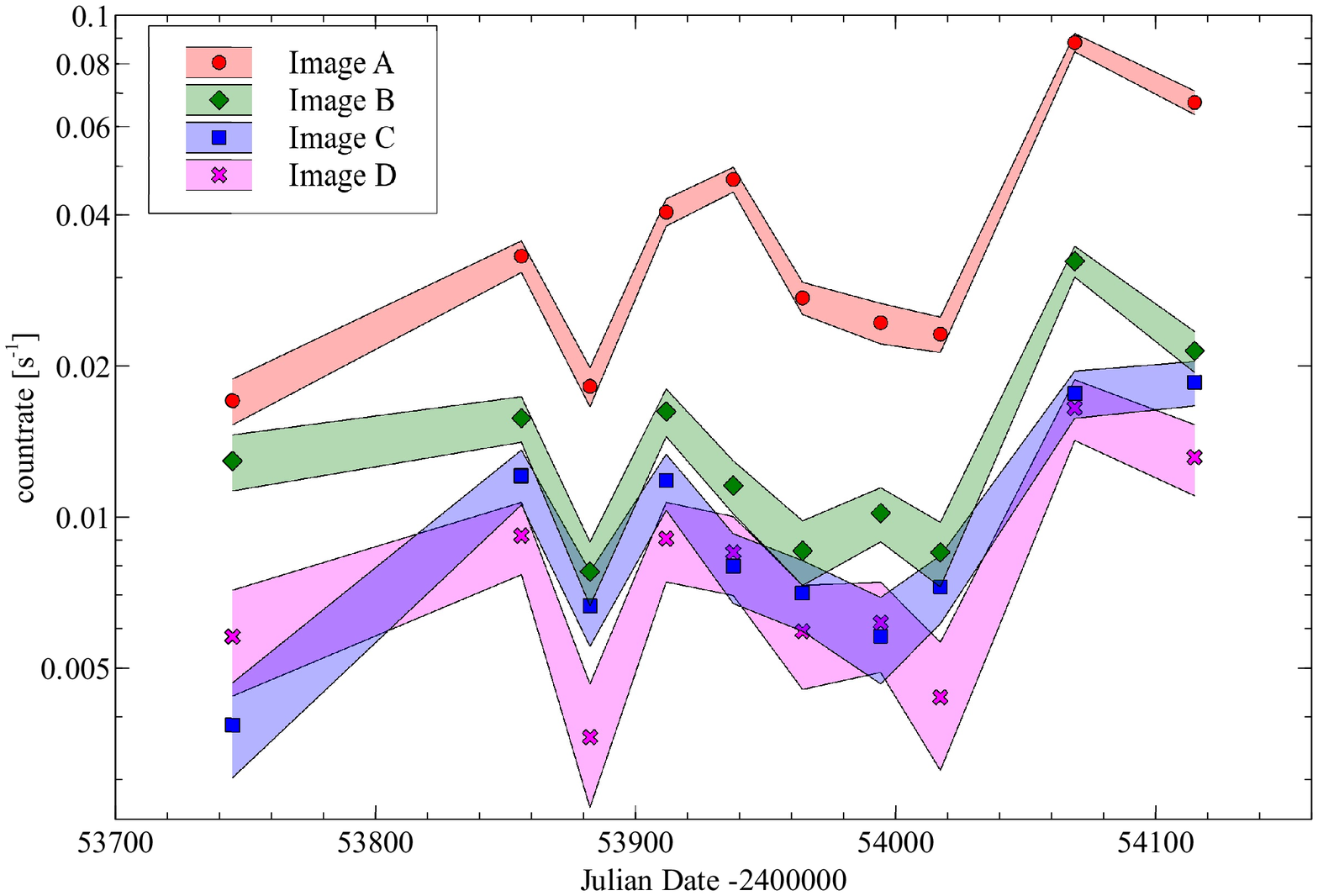}
	\caption[X-ray light curve of Q2237+0305 as obtained by using
          the C-statistic]{The final X-ray light curve of the four
          images of Q2237+0305 in the 0.5 keV - 8 keV band as obtained
          with the C-statistic based GALFIT. The figure shows the
          variations with time for all four lensed images. The
          datapoints and errorbars have been connected with shaded
          bands for each quasar image. A sudden increase of flux due
          to intrinsic variations in `the central engine' of the
          quasar can be seen for the last two epochs. The
          corresponding data can be found in Table
          \ref{cashresulttable}.}
\label{cashlightcurve}
  \end{minipage}
  \end{center}
\end{figure*}

The final fluxes for all ten observations and all four lensed quasar
images in the 0.5 keV - 8 keV band are shown in Table
\ref{cashresulttable}. The visualization of the data is shown in
Figure \ref{cashlightcurve} where we present the X-ray light curve
that shows the flux variations for the ten epochs spread over one year
of quasar image A (circle), B (diamond), C (square) and D (cross).

\begin{table*}
\centering
\begin{minipage}{140mm}
\centering
\caption[Fitting results obtained from $C$-statistic fitting]{X-ray
  fluxes of the four images of Q2237+0305 derived from the best-fit
  GALFIT runs. The flux for each observation and each lensed quasar
  image A, B, C and D in the 0.5 keV to 8 keV band is shown.\\}
\label{cashresulttable}
\begin{tabular}{@{}cccccc@{}}
\hline
  \textbf{Observation ID} & \textbf{Julian Date} & {\textbf{A}} & {\textbf{B}} & {\textbf{C}} & {\textbf{D}}\\
  \textbf{} & \textbf{} &  \multicolumn{4}{c}{count rate [$10^{-4}\cdot s^{-1}$]}\\
\hline
 6831 & 2453745.0 & $170.7\pm17.9$ & $129.4\pm16.5$ & $38.6\pm8.3$ & $57.8\pm13.8$ \\
 6832 & 2453856.0 & $331.4\pm23.9$ & $157.5\pm16.4$ & $120.1\pm15.1$ & $92.0\pm15.1$\\
 6833 & 2453882.4 & $182.3\pm16.4$ & $78.0\pm11.3$ &  $66.6\pm11.3$ & $36.5\pm10.1$ \\
 6834 & 2453911.7 & $405.7\pm25.2$ & $162.5\pm17.6$ & $118.4\pm15.1$ & $90.7\pm16.4$ \\
 6835 & 2453937.5 & $471.3\pm26.7$ & $115.6\pm14.0$ & $80.0\pm12.7$ &  $85.1\pm15.3$\\
 6836 & 2453964.2 & $273.7\pm20.2$ & $85.8\pm12.6$ & $70.6\pm11.35$ & $59.3\pm13.9$\\
 6837 & 2453994.2 & $244.2\pm22.6$ & $102.0\pm12.6$ & $57.9\pm11.3$ & $61.7\pm12.6$\\
 6838 & 2454017.2 & $231.7\pm18.8$ & $85.2\pm12.5$ & $72.6\pm11.3$ & $43.8\pm12.5$\\
 6839 & 2454068.8 & $882.8\pm36.8$ & $323.9\pm22.9$ & $176.5\pm19.1$ & $165.1\pm22.9$ \\
 6840 & 2454114.9 & $670.7\pm36.4$ & $214.4\pm20.1$ & $185.6\pm18.8$ & $131.6\pm21.3$ \\
\hline
\end{tabular}
\end{minipage}
\end{table*}

Although the light curve is only sparsely sampled it shows flux
variabilities in a wide range and for each of the four quasar
images. Most prominent are the variations in image A and the general
and very steep increase in the fluxes of all four images for the last
two epochs. As we are looking at images originating from a single
quasar, variations in all four images are due to processes in the
quasar itself because the time-delay is expected to be of the order of
a few hours only (see below). However, stars in the lensing galaxy may
also induce brightness variations through microlensing which hence are
different for each of the quasar images.  In the following we will
separate both types of variability.

\subsection{Microlensing}

Microlensing in the optical is a well known phenomenon in Q2237+0305
(e.g. recent studies including
\citet{Kochanek,timo_paper,Eigenbrod,Poindexter}). Thus we here
analyse the X-ray light curve to find evidence for X-ray microlensing
in `The Einstein Cross'.\\ If part of the flux variations are due to
microlensing, the pairwise flux ratios of the quasar images should not
be constant. We therefore have to consider that the flux $F$ of each
image $X$ at time $t$ is a result of different factors.

\begin{equation}
F^{X}(t)=\mu_{\mathrm{macro}}^{X}\cdot\mu_{\mathrm{micro}}^{X}(t)\cdot Q(t) 
\end{equation}
\\
Hence the brightness is affected by:

\begin{itemize}
 \item \textbf{$Q(t)$}: The time-dependent flux of the quasar as the source of the X-ray radiation. It is the same for each quasar image.
 \item \textbf{$\mu_{\mathrm{macro}}^{X}$}: The magnification factor
   that originates from the macro lens model and its local lensing
   parameters $\kappa$ and $\gamma$. It therefore depends on which
   quasar image we are looking at.
 \item \textbf{$\mu_{\mathrm{micro}}^{X}(t)$}: The variable magnification factor due to microlensing changes with time and is different for each quasar image. 
\end{itemize}

We do not need to correct for time-delays between the images as they
are negligible \citep{lensmodel_schneider, lensmodel_rix,
  timedelay_joachim, Robert_stronglens, photonindex,
  timedelay_koptelova}.  As we are interested in microlensing we form
the pairwise flux ratios between all quasar images $X$ and $Y$
($X,Y=1,..,4;\;X\neq Y$):

\begin{equation}
\dfrac{F^{X}(t)}{F^{Y}(t)}=\dfrac{\mu_{\mathrm{macro}}^{X}\cdot\mu_{\mathrm{micro}}^{X}(t)\cdot
  Q(t)}{\mu_{\mathrm{macro}}^{Y}\cdot\mu_{\mathrm{micro}}^{Y}(t)\cdot
  Q(t)}=\dfrac{\mu_{\mathrm{macro}}^{X}\cdot\mu_{\mathrm{micro}}^{X}(t)}{\mu_{\mathrm{macro}}^{Y}\cdot\mu_{\mathrm{micro}}^{Y}(t)}
\label{ratioequation}
\end{equation}

\begin{figure*}
     \begin{center}
\begin{minipage}{140mm}
     \includegraphics[scale=.85]{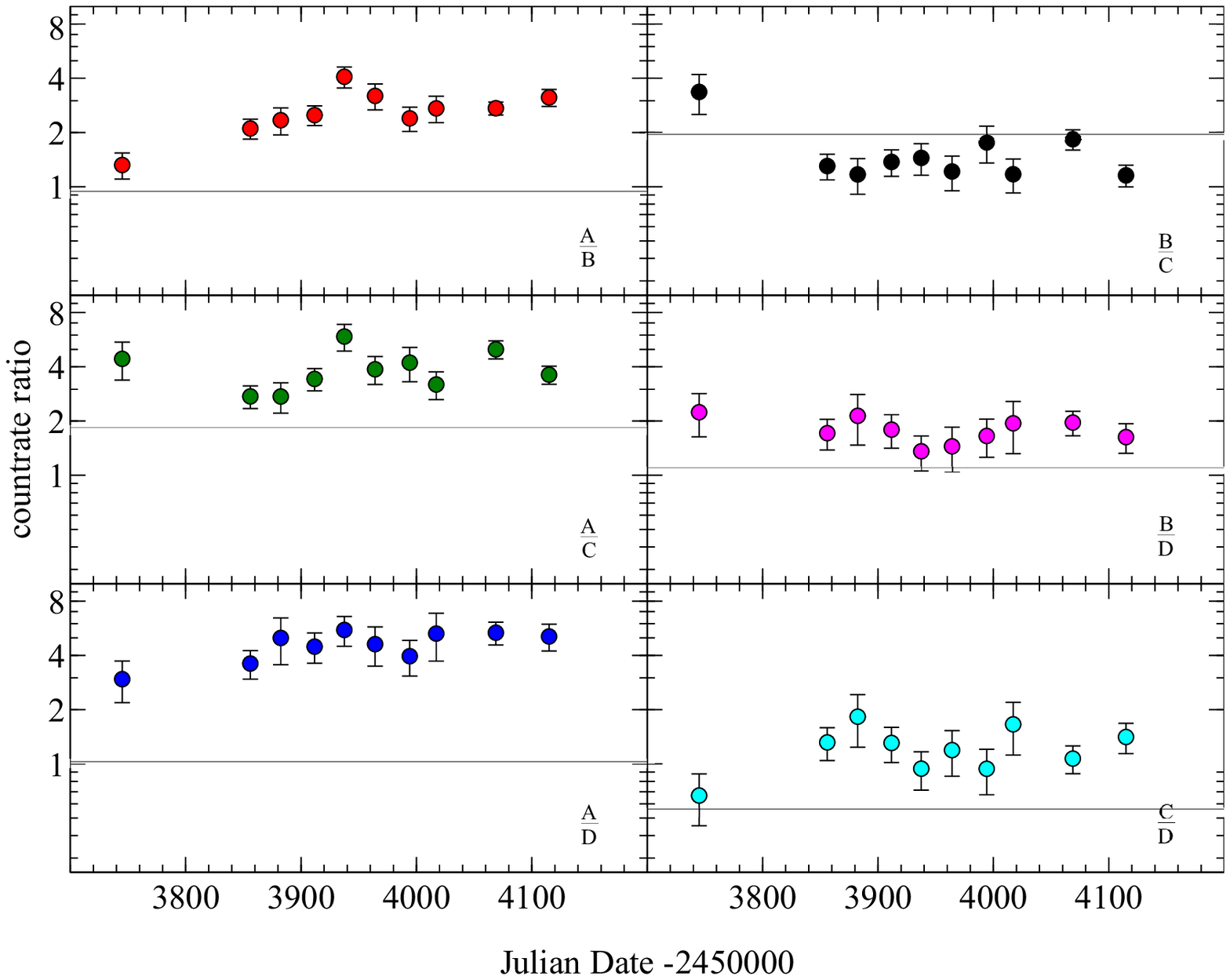}
      \caption[Pairwise count rate ratios of the four images corrected
        for macro magnification]{Pairwise X-ray count rate ratios of
        the four quasar images A through D of Q2237+0305 in order to
        identify microlensing. The ratios are calculated according to
        Equation \ref{ratioequation}. The horizontal lines mark the
        MIR flux ratios \citep{AgolMIR}, i.e. the macro magnification
        ratios.}
\label{ratiosixplot}
\end{minipage}
\end{center}
\end{figure*}

This way the intrinsic fluctuations of the quasar cancel out.  The
results are shown in Figure \ref{ratiosixplot} where we
logarithmically plotted the pairwise flux ratios of the quasar images
for every observation and according to equation
(\ref{ratioequation}). The difference to a light curve including the
macro magnification $\mu_{\mathrm{macro}}$ is just a constant offset
in the logarithmic plot and is of no importance for our
analysis. Additionally, we also plotted the MIR flux ratios
\citep{AgolMIR}, i.e. the macro magnification ratios, for each pair of
quasar images.

In Figure \ref{ratiosixplot} we see that the ratios $\frac{A}{B}$ and
$\frac{A}{C}$ show a similar variation signature (with the
  exception of the first data point), i.e. an increasing flux ratio
that has a maximum in the fith epoch and decreases afterwards. This behaviour is not mirrored by the ratio $\frac{B}{C}$ for these
epochs, which supports the view that the maximum in the ratios
$\frac{A}{B}$ and $\frac{A}{C}$ is caused by microlensing in image A.
The variations of ratios $\frac{B}{D}$ and $\frac{C}{D}$
are consistent with an independent low-amplitude microlensing signal in image D when compared with the ratio $\frac{B}{C}$.
However, the evidence for microlensing is most prominent in image A and the
ratio $\frac{A}{B}$ which is why we only focus on this quasar image
and image ratio, respectively.

The arguments given above and the fact that the correlated variations
especially in the ratio $\frac{A}{B}$ do not support random
fluctuations as the cause, leads to the conclusion that we have
observed microlensing induced variabilities. However, by comparing the observed
flux ratios with the macro magnification ratios, we find that all
quasar images may be affected by microlensing because their ratios are
not consistent with the MIR data \footnote{Note that the offset
  between the X-ray and IR fluxes can be slightly offset by
  differential absorption \citep{Eigenbrod_a}.}.  The flux ratios can
also be affected by substructures in the lensing halo, however, due to
the higher mass scale the time-scale of the variations induced by
substructures is much longer (due to the higher mass-scale) than the
time-scale considered in our analysis \citep{metcalf, Xu}.

\begin{figure*}
    \begin{center}
    \begin{minipage}{140mm}
    \centering
    \includegraphics[bb=0 415 426 842, scale=0.8]{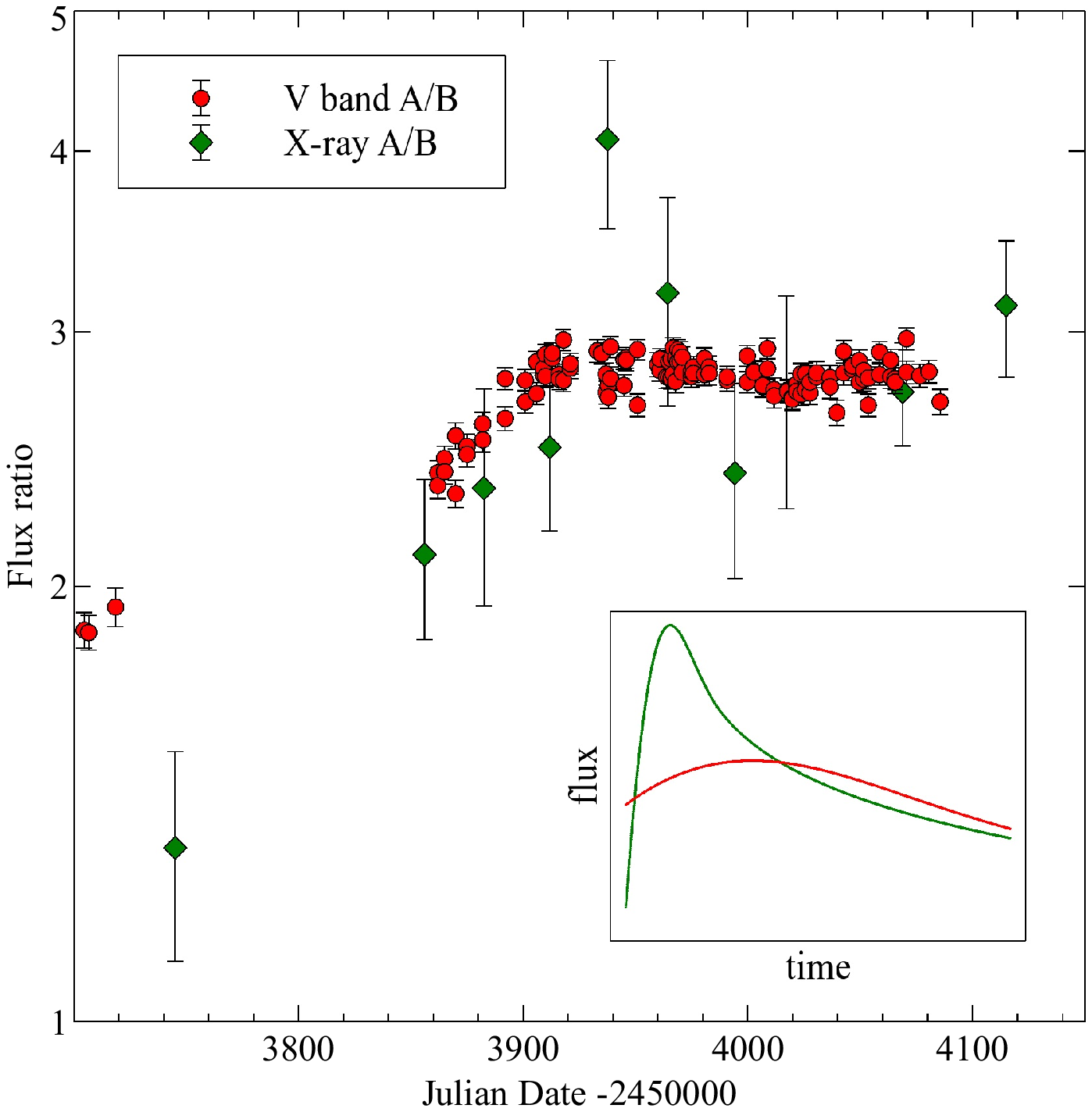}
    \caption[X-ray and optical flux ratios between image A and B]{Flux
      ratios of images A and B of Q2237+0305 in the X-ray regime in
      the 0.5 keV - 8 keV band (green diamonds) and the optical V-band
      (red circles) from January 2006 (2453745.0) to January 2007
      (2454114.9). We applied an additional systematic error of 0.03
      mag to the \textit{OGLE} data. The inset shows theoretical light
      curves for two Gaussian sources crossing a caustic, their radii
      differ by a factor of 6.7.}
\label{A-B_ogle_xray}
\end{minipage}
\end{center}
\end{figure*}

Focussing on the ratio of image A to image B and comparing it to OGLE
\citep{OGLE1,OGLE2,OGLE3} observations (V-band, restframe UV) obtained at the same
time, we find that the maximum as well as the subsequent flattening of the ratio in the X-ray regime
coincides with the optical observations. This is shown in Figure
\ref{A-B_ogle_xray} where the $\frac{A}{B}$ ratio in the X-rays is
overplotted with the OGLE data.  For the most prominent peak around
Julian date 2453937.5, the increase since Julian date 2453856.0 of the
X-ray ratio A/B is larger by a factor $\sim 1.6$ compared to the signal in
the optical. Furthermore, also the A-B difference light curve in the
UV \citep{Eigenbrod} for that period shows similar
characteristics. However, the events are less distinct at UV and
optical wavelengths which is exactly what we expect as the UV and the
optical emitting regions are thought to be larger than the X-ray
emitting region \citep{wampacz,rauch,photonindex,pooley,koch07}.
The inset in Figure \ref{A-B_ogle_xray} shows theoretical light curves for two Gaussian sources crossing a straight caustic. The radii are different by a factor of 6.7.

\subsection{Intrinsic variations}

The fact that there is strong evidence for microlensing in image A and
D does not explain the high flux rates for the last two epochs and the
flux rate drop for the third observation as the microlensing in image
A had its peak at JD 2453937.5 (July 2006). These fluctuations can
only be explained by intrinsic variations of the quasar itself. X-ray
variability on such short time scales of weeks indicates that the
emission generation takes place in a relatively compact region and
that the emission processes, that make quasars so luminous, are highly
variable. Such strong variations are not seen in the optical OGLE
light curve. Merely, the minimum at JD $\sim$ 2454010 can be seen.

\section{Summary and Outlook}

In this paper we have analysed archival \textit{Chandra} data of the
gravitationally lensed quasar Q2237+0305.  We compiled an X-ray light
curve for all four images. The data set comprises ten epochs ranging
from January 2006 to January 2007. Because of the blended nature of
the four images (see Figure \ref{6831_raw_labled}) it was necessary to
simultaneously fit appropriate PSFs to all four images in order to
obtain proper photometry. For this, we used a simulated PSF which
accounts for the optical properties of the \textit{Chandra}
observatory and the spectrum of the quasar X-ray emission. The fitting
was accomplished by a two-dimensional fitting algorithm and a grid
search by minimising Cash's C-statistic. Finally we analysed the light
curve and found evidence for microlensing variations in quasar image A
(see Figure \ref{ratiosixplot}). The X-ray microlensing signal in
image A coincides with the signal in the optical OGLE light
curve. Assuming that this parallel behaviour is caused by the same
process, i.e. one source that is microlensed, the amplitude of the
microlensing signal is a direct measure for the source size in the
particular wavelength regime \citep{wampacz,livingreview, Kochanek,
  timo_paper, Eigenbrod}. As the microlensing signal in the X-ray
regime is much more prominent than in the optical, this suggests that
the X-ray emission region is much smaller than the optical emission
region of the quasar \citep{pooley2, dai}. In a future paper (Zimmer
et al. 2010, \textit{in prep.}) we will make use of this effect to
measure the size of the X-ray and the optical emission region in
Q2237+0305.  While the time-delays in Q2237+0305 are negligible the
intrinsic variations have to be considered in the analysis. Figure
\ref{cashlightcurve} clearly shows how strong these variations are in
the X-ray regime. We do not find such strong fluctuations in the
optical. This indicates that the mechanism powering the quasar is
variable and leads to brightness variations by a factor of $\sim$ 4 on
time-scales of less than 30 days.\\

\section*{Acknowledgments}

We acknowledge the use of GALFIT 2.03 and thank the GALFIT team for
providing an adaptable GALFIT source code. We also like to acknowledge
the usage of the plotting package \textit{Veusz} and its lead author
Jeremy Sanders.

\appendix

\section[]{Cash's C-statistic}

In comparison to $\chi^{2}$-statistic Cash's C-statistic \citep{Cash}
is based on the assumption that each datapoint $y_{i}$ is a random
variable out of a Poissonian distribution with an expected value of
$\tilde{y_{i}}(z,a_{1},..,a_{M})$. Therefore, it has the probability:

\begin{equation}
p(y_{i})=\dfrac{\tilde{y_{i}}(z,a_{1},..,a_{M})^{y_{i}}\cdot e^{-\tilde{y_{i}}(z,a_{1},..,a_{M})}}{y_{i}!}
\end{equation}

So the probability of the whole data set is the product over all datapoint probabilities $p(y_{i})$.

\begin{equation}
P=\displaystyle\prod_{i=1}^Np_{i}(y_{i})
\label{datasetequation}
\end{equation}

To maximize the data set probability by estimating the model parameters $a_{1},..,a_{M}$ one can also minimize  $-2\cdot\log(P)$: 

\begin{equation}
  \begin{split}
  -2\cdot log(P)=-2\cdot\displaystyle\sum_{i=1}^{N}&y_{i}\cdot\log(\tilde{y_{i}}(z,a_{1},..,a_{M}))-\\&\tilde{y_{i}}(z,a_{1},..,a_{M})-\log(y_{i}!)
   \end{split}
\end{equation}

leading to the definition of $C$:

\begin{equation}
C=-2\cdot\displaystyle\sum_{i=1}^{N}y_{i}\cdot \log(\tilde{y_{i}}(z,a_{1},..,a_{M}))-\tilde{y_{i}}(z,a_{1},..,a_{M})
\end{equation}

The term $-\log(y_{i}!)$ can be left out as it does not depend on the
model parameters and is therefore just a constant offset. For our
implementation in GALFIT we chose a slightly modified form as
implemented in Xspec \citep{xspec}:

\begin{equation}
 \begin{split}
    C=-2\cdot\displaystyle\sum_{i=1}^{N} y_{i}\cdot& \log(\tilde{y_{i}}(z,a_{1},..,a_{M}))-\tilde{y_{i}}(z,a_{1},..,a_{M})\\&  +y_{i}-y_{i}\cdot \log(y_{i})
  \end{split}
\end{equation}

This translates to the following function to be minimized by GALFIT:

\begin{equation}
\begin{split}
C=-2\cdot\displaystyle\sum_{x=1}^{nx}\displaystyle\sum_{y=1}^{ny}&\mathrm{flux}_{x,y}\cdot
\log(\mathrm{model}_{x,y})-\mathrm{model}_{x,y}+\\&\mathrm{flux}_{x,y}-\mathrm{flux}_{x,y}\cdot
\log(\mathrm{flux}_{x,y})
\end{split}
\end{equation}

\bsp

\label{lastpage}

\end{document}